# A New cryptanalysis model based on random and quantum walks

Ahmed DRISSI
idrissi2006@yahoo.fr
National School for Applied Sciences
AbdelMalek Essaadi University
Tangier, Morocco

Abstract
Randomness plays a key role in the design of attacks on cryptographic systems and cyber security algorithms in general. Random walks and quantum walks are powerful tools for mastering random phenomena. In this article, I propose a probabilistic attack model of a cryptographic system. This model is based on a random or quantum walk with the space of states being a space containing the secret to be revealed. This space can be a subspace of keys, plain texts or cipher texts.

## I. Introduction

The principle behind the majority of probabilistic attacks is to browse a space containing a secret to be revealed (space of keys, space of encrypted or clear texts, etc.) randomly in order to reach the element sought. The mathematical and statistical study of this random walk provides us with sufficient tools for the design of a general probabilistic attack on a cryptographic system.

A random walk is a stochastic process that describes a path that includes a succession of random steps in a mathematical space. There are two categories, one classical and the other quantum. The main difference between classical random walks and quantum random walks is that quantum random walks do not converge to certain limiting distributions. The time complexity is lower in the quantum case than in the classical case. Algorithms based on quantum random walks can provide an exponential speedup compared to any classical algorithm. So they are often used to speed up classical algorithms. In 2001,[4] proposes a scheme to implement the quantum random walk on a line and on a circle in quantum computer. [5] use quantum walks to construct a new quantum algorithm for element distinctness and its generalization. In 2020 [1] the authors provides a comprehensive review of classical random walks and quantum walks. And they compare the algorithms based on quantum walks and classical random walks from the perspective of time complexity.

The following section will be devoted to the mathematical study of random and quantum walks. The richness of mathematical and algorithmic properties shows that the random walk is a good candidate for generalizing probabilistic attacks. A general model of attacks based on random and quantum walks is presented in the third paragraph. We end with a short conclusion.

## II. Mathematical study of random walks

Random walks are just homogeneous Markov chains. However, homogeneous Markov chains are rich in mathematical properties, which provides us with fairly powerful mathematical tools, useful for designing algorithms in different contexts. In this paragraph I present the mathematical and statistical properties of random walks and quantum walks useful for any subsequent design. Algorithm designers can rely on these properties according to their purposes.

### 1. The classical random walk on Z

Mathematically, it is a sequence of random variables$(S_n)_{n\in N}$ , with values in $Z$ defined by the abscissa of a particle moving by a step on Z to the right with probability p and to the left with probability $1-p$. ($p(X_i=1)=p$ and $p(X_i=-1)=1-p$) with $X_i$ : jump number $i$.

The abscissa of the particle, after n jumps, will be modeled by the random variable $S_n=\sum_{i=1}^{n}X_i$. It is easy to notice that $(X_i)_{i\in N}$ is a sequence of independent and identically distributed variables. $S_n$ is said to be a random walk on Z. in what follows we restrict ourselves to $p=\frac{1}{2}$ .

Lemma 1

The distribution of the random walk $S_n$ on Z is given by

$$p(S_n=i)=\begin{cases} C_n^{\frac{n+i}{2}}\left(\frac{1}{2}\right)^n & \text{if } n \text{ and } i \text{ have the same parity} \\ 0 & otherwise \end{cases}$$

.

Proof

let's calculate $p(S_n=k)$, we make the following transformation : $B_i=\frac{X_i+1}{2}$, $p(X_i=1)=p(B_i=1)=p$ and $p(X_i=-1)=p(B_i=0)=1-p$. We have $B_i \sim B(p)$ then $Y_n=\sum_{i=1}^{n}B_i \sim B(n,p)$.

$Y_n=\sum_{i=1}^{n}B_i=\sum_{i=1}^{n}\frac{X_i+1}{2}=\frac{1}{2}(S_n+n)$

and we have $Y_n \epsilon N$ and $S_n \epsilon Z$ then if $S_n+n$ is even we will have a $Y_n \epsilon N$ Otherwise $Y_n$ cannot take on any value (i.e $p(Y_n=odd)=0$).

$p(S_n=i)=p(2Y_n-n=i)=$
$p\left(Y_n=\frac{n+i}{2}\right)=C_n^{\frac{n+i}{2}}p^{\frac{n+i}{2}}(1-p)^{\frac{n-i}{2}}$ if $\frac{n+i}{2}\in N$

$p(S_n=i)=0$ if $\frac{n+i}{2}\notin N$

Lemma 2

$S_n$ follows a binomial law, its expected value $E(S_n)=0$ and its variance is n (standard deviation $\sqrt{n}$)

Lemma 3

The random walk is a homogeneous Markov chain.

Proof

The random walk verifies the Markov property, the next abscissa of the particle depends on its past only through its present, indeed the transition probabilities

$$p_{ij}=p(S_{n+1}=j \;/\; S_n=i) = \frac{p(S_{n+1}=j\;,S_n=i)}{p(S_n=i)} = \frac{p(X_{n+1}+i=j\,)}{p(S_n=i)} = \frac{p(X_{n+1}=j-i\,)}{p(S_n=i)}$$

$$p_{ij}=\begin{cases} \dfrac{\frac{1}{2}C_n^{\frac{n+i}{2}}\left(\frac{1}{2}\right)^n}{C_n^{\frac{n+i}{2}}\left(\frac{1}{2}\right)^n}=\frac{1}{2} & if\ j=i\pm 1 \\ 0 & otherwise \end{cases} .$$

(if $j-i=\pm 1$ , j is neighbor to i)

Transition Matrix

$$P=\begin{bmatrix} & \ddots & \ddots & \ddots & \ddots & & & \\ & 0 & \frac{1}{2} & 0 & \frac{1}{2} & 0 & & \\ & & 0 & \frac{1}{2} & 0 & \frac{1}{2} & 0 & \\ & & & 0 & \frac{1}{2} & \ddots & \ddots & 0 \\ & & & & \ddots & \ddots & & \end{bmatrix}$$

Lemma 4

$$(p(S_0=1),p(S_0=2),\dots,p(S_0=k),\dots)P = (p(S_0=1),p(S_0=2),\dots,p(S_0=k),\dots)$$

$$(p(S_0=1),p(S_0=2),\dots,p(S_0=k),\dots)P^n = (p(S_n=1),p(S_n=2),\dots,p(S_n=k),\dots)$$

2. Random walk on a set

Given the set N (of a cardinal n) of which a random walker walks. Let the subset M of a cardinal m marked elements (having particularities compared to the other elements of N). Assuming that the walk distribution is uniform, the stochastic transition matrix P will be :

$$P_{ij} = \begin{cases} \frac{1}{n-1} \text{ if } i \neq j \text{ and } i \notin M \\ 0 \text{ if } i = j \text{ and } i \notin M \\ 1 \text{ if } i = j \text{ and } i \in M \\ 0 \text{ if } i \neq j \text{ and } i \in M \end{cases}$$

By rearranging the elements of N and index them as follows $\{x_1..x_m\} = M \subset N = \{x_1..x_m, x_{m+1} ... x_n\}$, the transition matrix becomes $P = \begin{pmatrix} I & 0 \\ P_1 & Q \end{pmatrix}$

i.e

$Q = \frac{1}{n-1}\begin{pmatrix} 0 & 1 & \cdots & 1 \\ 1 & 0 & \ddots & \vdots \\ \vdots & \ddots & 0 & 1 \\ 1 & \ldots & 1 & 0 \end{pmatrix}$ : The stochastic transition matrix from unmarked elements to unmarked elements (between transient states).

$P_1 = \frac{1}{n-1}\begin{pmatrix} 1 & \ldots & \cdots & 1 \\ \vdots & \cdots & \ldots & \vdots \\ 1 & \ldots & \ldots & 1 \end{pmatrix}$ : The stochastic transition matrix from unmarked elements to marked elements (recurrent states).

$I_m$ : the identity matrix.

Lemma 5

Q is an invertible matrix.

Let us note

$p_{ij}$ : the probability, starting from i, to reach j.

$p_{ij}^{(n)}$ : The probability, starting from i, to reach j after n experiences (or transitions).

$f_{ij}$ : The probability, starting from i, , to reach j (in a finite number of transitions).

$f_{ij}^{(k)}$ : The probability, starting from i, to reach j for the first time in k transitions.

By referring to [10], it is easy to extract the following mathematical results:

Lemma 6

The random walk on Z is an irreducible homogeneous Markov chain.

Lemma 7

The random walk on N is a homogeneous Markov chain with N\M is the set of transient states and for all $x \in M$,$\{x\}$ is a recurrent class.

Theorem 1

$$\text{For all } j \in M, \begin{bmatrix} f_{1j} \\ \vdots \\ f_{ij} \\ \vdots \\ f_{(n-m)j} \end{bmatrix} = Q \begin{bmatrix} f_{1j} \\ \vdots \\ f_{ij} \\ \vdots \\ f_{(n-m)j} \end{bmatrix} + \begin{bmatrix} p_{1j} \\ \vdots \\ p_{ij} \\ \vdots \\ p_{(n-m)j} \end{bmatrix}$$

Theorem 2

Let $N_i$ the number of transition until the absorption to a recurrent class, starting from i and $n_i = E[N_i]$ (the average number). We will $(I - Q)\begin{bmatrix} n_1 \\ \vdots \\ n_{(n-m)} \end{bmatrix} = \begin{bmatrix} 1 \\ \vdots \\ 1 \end{bmatrix}$.

Theorem 3

If $m = 1$ (that's to say M contains only one element) the system $\begin{cases} P\pi = \pi \\ \sum_{i=1}^{n} \pi_j = 1 \end{cases}$ admits a unique solution then $\lim_{n\to\infty} p_{ij}^{(n)} = \pi_j, \forall i, j \in N$.

## 3. The quantum walk

Appeared in 2003, the quantum walk generalizes the discrete random walk, to the quantum case, using reasoning by analogy. Instead of a walker, we consider a quantum state $||\Psi_t>$ which depends on both the position and an internal degree of freedom. Quantum walk models that can be used to perform the same tasks as classical walks. The conditional shift operator for the quantum walk on the line is given by $S = |0><0| \otimes \sum_i |i+1><i| + |1><1| \otimes \sum_i |i-1><i|$ That is, the particle jumps to the right if it has spin up and to the left if it has spin down. Explicitly, the conditional shift operator acts on the product states according to $S(|0> \otimes |i>) = |0> \otimes |i+1>$ and $S(|1>\otimes|i>) = |1>\otimes|i-1>$, which means the particle jumps right if it spins up and left if its spins down. H is a unitary transformation that rotates the spin. The discrete-time quantum walk is defined by two operators: H and S. Each step is determined by the application of $U = S(H\otimes I)$, $U|k>|j> = \sum_{k=0}^{1}\frac{(-1)^k}{\sqrt{2}}|k>|j+(-1)^k>$. The quantum walk after n steps is $U^n$.

let's write $U|k>|j>$ in matrix form and deduce the form of $U^n|k>|j>$. How should we proceed ? Use U or $U = S(H\otimes I)$. We start with the following example:

**Example**

Consider a quantum walk starting with the state $|\Psi_0(0)>=|0>|0>$, after 4 steps :

$|\Psi_1> = \frac{(|0>|1>+|1>|-1>)}{\sqrt{2}}$,

$|\Psi_2> = \frac{(|0>|2>+|1>|0>+|0>|0>-|1>|-2>)}{2}$

$|\Psi_3> = \frac{(|0>|3>+|1>|1>+2|0>|1>-|0>|-1>+|1>|-3>)}{2\sqrt{2}}$,

$|\Psi_4> = \frac{(|0>|4>+|1>|2>+3|0>|2>+|1>|0>-|0>|0>-|1>|-2>+|0>|-2>-|1>|-4>)}{4}$

,

Whose distribution is as follows:

| | -4 | -3 | -2 | -1 | 0 | 1 | 2 | 3 | 4 |
|---|---|---|---|---|---|---|---|---|---|
| 0 | | | | | 1 | | | | |
| 1 | | | | 1/2 | | 1/2 | | | |
| 2 | | | 1/4 | | ½ | | ¼ | | |
| 3 | | 1/8 | | 1/8 | | 5/8 | | 1/8 | |
| 4 | 1/16 | | 2/16 | | 2/16 | | 10/16 | | 1/16 |

We can see that the distribution becomes more and more skewed to the right, whereas in the classical case the distribution will be symmetric around the starting position.

Now we assume that the walk begins in the state $|\Psi_0>=|1>|0>$,

$|\Psi_1> = \frac{(|0>|1>+|1>|-1>)}{\sqrt{2}}$,

$|\Psi_2> = \frac{(-|0>|2>-|1>|0>+|0>|0>-|1>|-2>)}{2}$,

$|\Psi_3> = \frac{(-|0>|3>-|1>|1>+2|1>|-1>-|0>|-1>+|1>|-3>)}{2\sqrt{2}}$,

$|\Psi_4> = \frac{(-|0>|4>-|1>|2>-|0>|2>+|1>|0>+|0>|0>-3|1>|-2>+|0>|-2>-|1>|-4>)}{4}$,

The distribution of which is as follows:

| | -4 | -3 | -2 | -1 | 0 | 1 | 2 | 3 |
|---|---|---|---|---|---|---|---|---|
| 0 | | | | | 1 | | | |
| 1 | | | | ½ | | ½ | | |
| 2 | | | ¼ | | 1/2 | | 1/4 | |
| 3 | | 1/8 | | 5/8 | | 1/8 | | 1/8 |
| 4 | 1/16 | | 10/16 | | 2/16 | | 2/16 | 1/16 |

The distribution given by this step is the symmetry of the first. To generate a symmetric distribution, we consider the initial state $|\Psi(0)>=\frac{1}{\sqrt{2}}(|0>-i|1>)|0>$.

Theorem 4 [3]

$$U^n|0>|0>=\sum_{i=}^{n} u_i|i>|0>+\sum_{i=}^{n} v_i|i>|1>$$

$$u_i = \begin{cases} \frac{1}{\sqrt{2^n}} & if \quad i = n \\ \frac{1}{\sqrt{2^n}}\sum_{k=1} C_{\frac{n-i}{2}-1}^{k-1} C_{\frac{n+i}{2}}^{k}(-1)^{\frac{n-i}{2}-k} & if \quad i < n \end{cases}$$

$$v_i = \frac{1}{\sqrt{2^n}}\sum_{k=0} C_{\frac{n-i}{2}-1}^{k} C_{\frac{n+i}{2}}^{k}(-1)^{\frac{n-i}{2}-k-1}$$

$u_i = v_i = 0\ if\ \ n+i\ \ odd\ \ or\ |i| > n$

We deduce from this theorem the following lemma which gives us an explicit form of probability of being in a position i after n transitions.

Theorem 5

$p(S_n = i) = 0$ if $n+i$ is odd or $|i| > n$

$p(S_n = i) = |u_i|^2 +$

$|v_i|^2 = \frac{\left|\sum_{k=1} C_{\frac{n-i}{2}-1}^{k-1} C_{\frac{n+i}{2}}^{k}(-1)^{\frac{n-i}{2}-k}\right|^2 + \left|\sum_{k=0} C_{\frac{n-i}{2}-1}^{k} C_{\frac{n+i}{2}}^{k}(-1)^{\frac{n-i}{2}-k-1}\right|^2}{2^n}$ if $i < n$

$p(S_n = \mp n) = \frac{1}{2^n}$ .

## 4. Grover's Algorithm

Grover's algorithm is an algorithm that aims to solve the following problem: Find an element $x_0 \in \{1 \dots N\}$ having a particular criterion.

Instance

Given the set $\{1 \dots N\}$ and a function $f: \{1 \dots N\} \to \{0,1\}$ ; $x \to 1$ if $x = x_0$ and 0 otherwise.

Find $x_0$ .

Procedure

- Preparing a superposed state (the probability of measuring $x_0$ is $\frac{1}{2^n}$). we need $2^n$ times to guess the

correct iteration. ($2^n \leq N < 2^n + 1$).

- The algorithm proceeds in such a way that a black box inverts the phase of the state verifying the specific criterion $x_0$.
- Repeating the procedure increases the probability of measuring $x_0$.

Grover's search can be thought of as a quantum walk algorithm, which brings us to Ambainis' quantum walk-based algorithm from [1] for the element distinction problem, which gives a speedup over classical and other non-step-based quantum algorithms. Grover's search algorithm requires $\sqrt{N}$ quantum queries.

## III. Our cryptanalysis model

Probabilistic attacks are algorithms that aim to reveal a certain secret, with a certain probability. The attacker exploits the information he has to increase this probability. On the other hand, we can follow the same way in order to avoid any temptation to attack.

Both classical random walks and quantum random walks can be used to calculate the proximity between nodes and extract topology in a network. Indeed, walking on a set K, starting from i ∈ K and aiming to achieve j ∈ K (desired), We aim to answer the following three main questions :

- How many steps are expected for the walk to visit all elements j of K, in the case n = | K | is finished ?
- Given a point j different from i, what is the expected number of steps to reach j from $i$ ?
- If you stop walking after a given number n steps, what is the probability $p_{ij}^n$ to end at the knot j ? How does it behave when n get big? ($\lim_{n\to\infty} p_{ij}^n$).

On the other hand, in [12] the authors introduced an extended random walk model, in a social network, to describe the browsing behavior of users and establish the legitimate pattern of browsing sequences. For each incoming browser, its page request sequence is observed and the next page request sequence is predicted based on the random walk model. The similarity between the predicted and observed page request sequence is used as a criterion to measure the legality of the user, then the attacker would be detected based on this. Another application is given by Kaplan, M [11], who devised quantum attacks against iterated block ciphers. The literature includes several applications of random and quantum walks in cryptanalysis. We are interested here in giving a general model. In mathematical notation a cryptographic system can be defined as a tuple $(P, C, K, E, D)$ with

$P$: finite set of plaintext

$C$ : finite set of ciphertext

$K$ :set of possible keys k.

$E$ : cipher function set $e_k: P \to C, x \to y = e_k(x)$

$D$ : decryption function set $d_k: C \to P, y \to x = d_k(y)$

In general, attack scenarios are:

- Ciphertext-only attack

The attacker only has the ciphertext $y \in C$ and look for the key k or at least the corresponding plain text x.

- Known plaintext attack

The attacker owns the plaintext x and the corresponding ciphertext y. He is looking for the key k checking $y = e_k(x)$.

- The chosen plaintext attack

The attacker has gained temporary access to the encryption machine so he can choose a plaintext string x and construct the corresponding ciphertext string y.

- The chosen ciphertext attack

The attacker has gained temporary access to the decryption machine, so he can choose a ciphertexty and reconstruct the corresponding plaintext $x = d_k(y)$. He is looking for the key k.

The different attack models depend on the information available to the attacker.

Kerckhoff's principle [13] expresses that the security of a cryptosystem is based on the secrecy of the key and assumes that the two functions $e_k$ and $d_k$ are publicly known. So the main objective of cryptanalysis is to find the key, but in some cases the ambition of the attacker is limited to finding just a plaintext or a link between a limited number of plaintexts. In general, the problem and the general form of attack can be summarized as follows : Given a set S (containing the secret $x \in S$ to reveal). Find x.

Instance

Choose a finished set A of S ($A \subset S$).

Choose a point $x_0 \in A$

Walk randomly from $x_0$ vers x.

General principle

The probability of success increases if the size of S decreases what motivates the attacker to locate the key in a part of A as small as possible.

Exploiting the information available, and based on the mathematical and algorithmic properties detailed in the two previous paragraphs, the designer of a cryptanalysis makes the following decisions:

Choose A verifying that the probability that $x_0 \in A$ is closer to 1 and at least greater than $\frac{1}{2}$. ($\frac{1}{2} < p(x \in A) < 1$).

Minimize distance $d(x, x_0)$ between the unknown secret and the initialization poin $x_0$.

Minimize the number of steps n such that the probability $p = p_{x_0 x}^{(n)}$ to reach x starting from $x_0$ in n transitions or not, closer to 1.

Iterate this random walk at least $\frac{1}{p}$ times.

## IV. Conclusion

In this article, I tried to give a general approach of probabilistic attack of a cryptographic system based on the particular properties of random and quantum walks. For any cryptanalysis operation, the attack designer can use this model and adapt it to his particularities and optimize it using the mathematical results thus obtained. A work of analysis and deep simulation of each scenario apart remains a project for the future.